Spinel ferrites: old materials bring new opportunities for spintronics

Ulrike Lüders[1,2], Agnès Barthélémy[3]*, Manuel Bibes[4], Karim Bouzehouane[3], Stéphane Fusil[5], Eric Jacquet[3], Jean-Pierre Contour[3], Jean-François Bobo[2], Josep Fontcuberta[1] and Albert Fert[3]

[1] Institut de Ciència de Materials de Barcelona, CSIC, Campus de la Universitat Autònoma de Barcelona, 08193 Bellaterra, Catalunya, Spain

[2] Lab. de Nanomagnétisme pour l'Hyperfréquence CNRS-ONERA, BP4025, 2 avenue Edouard Belin, 31055 Toulouse, France

[3] Unité Mixte de Physique CNRS-Thales, Domaine de Corbeville, 91404 Orsay, France

[4] Institut d'Electronique Fondamentale, Université Paris-Sud, 91405 Orsay, France

[5] Université d'Evry, Bât. des Sciences, rue du père Jarlan, 91025 Evry, France

Abstract

Over the past few years, intensive studies of ultrathin epitaxial films of perovskite oxides have often revealed exciting properties like giant magnetoresistive tunnelling[1] and electric field effects[2]. Spinel oxides appear as even more versatile due to their more complex structure and the resulting many degrees of freedom. Here we show that the epitaxial growth of nanometric $NiFe_2O_4$ films onto perovskite substrates allows the stabilization of novel ferrite phases with properties dramatically differing from bulk ones. Indeed, $NiFe_2O_4$ films few nanometres thick have a saturation magnetization at least twice that of the bulk compound and their resistivity can be tuned by orders of magnitude, depending on the growth conditions. By integrating such thin $NiFe_2O_4$ layers into spin-dependent tunnelling heterostructures, we demonstrate that this versatile material can be useful for spintronics, either as a **conductive** electrode in **magnetic tunnel junctions** or as a spin-filtering **insulating** barrier in the little explored type of tunnel junction called **spin-filter**. Our findings are thus opening the way for the realisation of monolithic spintronics architectures integrating several layers of a single material, where the layers are functionalised in a controlled manner.

* : Fax : +33-1-69-33-07-40 ; email : agnes.barthelemy@thalesgroup.com

The search for novel functional materials is an extremely active field and to this purpose transition-metal oxides have been under focus for over 20 years now. Their properties mainly arise from the interaction between the transition metal and oxygen ions and its sensitivity to the bond length and angles[3]. Consequently, a large number of materials showing different physical effects can be found within one structural family. Perovskites are the most popular as this crystal structure is shared by high $T_C$ superconductors, ferroelectrics, half-metallic ferromagnets, etc. This structural compatibility allows one to combine thin layers of very different materials to design multifunctional epitaxial heterostructures. Furthermore, the strong sensitivity of the physical properties to structural modifications often reveals unexpected behaviour in strained thin films[4,5] or at the interface between two adjacent layers[6].

Besides perovskites, spinel oxides are also very attractive as they can be half-metallic (as $Fe_3O_4$ [7]), ferrimagnetic insulators (like most spinel ferrites [8]), transparent conductors (as $Cd_2SnO_4$ [9]), superconductors ($LiTi_2O_4$ [10]) or heavy-fermion compounds ($LiV_2O_4$ [11]). Furthermore, spinel is a complex crystal structure, with many degrees of freedom available to engineer physical properties. Yet, their study in thin film form has not been so intensive[12] and much remains to be learnt concerning their properties in low dimensions and their potential in heteroepitaxial architectures.

In this paper, we report on the growth and properties of ferrimagnetic $NiFe_2O_4$ thin (3-12 nm) films onto perovskites $SrTiO_3$ (STO) substrates, layers of the ferromagnetic metallic oxide $La_{2/3}Sr_{1/3}MnO_3$ (LSMO), or LSMO/STO bilayers. These nanometric films have a considerably enhanced magnetic moment, compared to bulk, and their electronic properties can be tuned from **conductive** to **insulating** by changing the growth conditions. We have integrated these layers into epitaxial heterostructures to probe their potential for spintronics. We find that both types of NiFe$_2$O$_4$ produce spin-polarized currents but via two different mechanisms that might be combined to fabricate novel spintronics devices.

NiFe$_2$O$_4$ (NFO) films have been grown by target-facing-target RF sputtering onto either STO(001) substrates or on LSMO/STO or LSMO templates (see Experimental). The growth atmosphere consisted either of pure Ar or of Ar/O$_2$ mixture. After growth, the films were cooled to room temperature in the growth atmosphere. Reflection high-energy electron diffraction (RHEED) patterns recorded before and after film deposition (figure 1a) indicate epitaxial growth of the spinel material onto the perovskite for both Ar and Ar/O$_2$ conditions. The pattern consists of organized spots for films grown in pure Ar and of streaks for Ar/O$_2$. This indicates that the growth mode is modified from three-dimensional to two-dimensional when O$_2$ is incorporated. Consequently, the surface of Ar/O$_2$ films is smoother, as shown in the atomic force microscope (AFM) images of figure 1c. High-resolution X-ray diffraction 2θ-ω (shown for two 12 nm films in figure 1b) and ϕ scans[13] confirm that the spinel material grows cube-on-cube onto the perovskite, with no indication of parasite phases. The Ni/Fe ratio was estimated to 0.5±0.05 from X-ray photoemission spectroscopy[13] and no interdiffusion between the film and the template was found from transition electron microscopy and electron energy loss spectroscopy [14].

Magnetic hysteresis cycles for two 6 nm films, grown in Ar or in Ar/O$_2$, are shown in figure 2. For both samples, the magnetization at large field (10kOe) is in the 600 emu.cm$^{-3}$ range, which is twice larger than the saturation magnetization of bulk NFO. This comes from the possibility of atomic rearrangements in the spinel structure. Bulk NFO is an inverse spinel with Fe$^{3+}$ ions in sublattice A, ordered Ni$^{2+}$ and Fe$^{3+}$ ions in sublattice B and an antiferromagnetic coupling between the two sublattices. With this ordering, the moments of the Fe ions in A and B cancel out, which results in a saturation magnetization of only 2$\mu_B$/formula unit i.e. 300 emu.cm$^{-3}$. If Ni ions from the B sublattice replace Fe ions in the A sublattice, and vice-versa, the moment of the Fe ions no longer compensate each other and the magnetization can increase dramatically. As our films are single-phase with the correct Ni/Fe ratio, the enhanced magnetic moment is likely to originate from cationic inversion. An enhanced magnetic moment due to cationic inversion has already been observed at the surface of spinel nanoparticles[15]. The Curie temperature of the NFO films is well beyond room temperature, as shown in the inset of figure 2.

While the presence of oxygen in the growth atmosphere does not modify substantially the magnetic response, it has a dramatic influence on the transport properties. Films grown in Ar/O$_2$ are insulating with a room-temperature resistivity exceeding at least $\rho \approx 200$ $\Omega$.cm (that is in the range found for bulk[8]) while films grown in pure Ar have a room-temperature conductivity three orders of magnitude larger ($\rho \approx 100$ m$\Omega$.cm). We note that this value is similar to that reported for conducting Fe$_3$O$_4$ films[16] of similar thickness. Moreover the temperature dependence $\rho(T)$ of a 12 nm NFO film grown in Ar (shown in the inset of Figure 2) also bears resemblance with magnetite. Ultraviolet spectroscopy[13] experiments evidence the presence of a finite density of states at the Fermi edge in these films, thus confirming its quasi-metallic character. This conductive behaviour may be related to the presence of mixed-valence Fe$^{2+/3+}$ ions induced by oxygen vacancies in the B sublattice and the subsequent charge delocalization. We have verified that the respective insulating and conductive properties of these films remain stable after one year of storage at room temperature. It thus appear that the electrical properties of NFO films can be controllably tuned by the growth conditions. NFO presents the additional advantage of being ferrimagnetic, which makes both conductive and insulating phases useful for spintronics, as will be shown later on.

The transport properties of these two types of NFO have been further examined by imaging 3 to 12 nm-thick layers grown onto LSMO/STO bilayers with a conductive-tip AFM[17] (CTAFM). In this type of experiments, the resistance between the bottom electrode (here LSMO) and the tip (in contact with the sample surface) is measured as the sample is scanned, so that a resistance map is collected. Some of these maps are shown in figure 3. Taking the resistance map of a LSMO/STO(0.8nm) bilayer as a reference, one can clearly see that, for a NFO thickness of 5 nm, the average resistance is roughly the same when the NFO film has been grown in pure Ar. In contrast, an increase of about 3 orders of magnitude is observed for a NFO film grown in $Ar/O_2$. The average resistance for these two types of NFO is plotted vs NFO thickness in figure 3b. As expected for conductive NFO, the resistance does not depend on thickness, while an exponential increase is obtained for insulating NFO, which indicates that transport occurs by tunnelling through this type of NFO layer grown in $Ar/O_2$.

To check the potential of **conductive** NFO for spintronics, LSMO(35nm)/STO(0.8nm)/ NFO(3nm) **magnetic tunnel junctions** have been prepared using CTAFM lithography[18]. In these structures, two ferromagnetic conducting electrodes, LSMO and NFO, are separated by a thin insulating STO barrier (see figure 4a). Figure 4c shows the dependence of the resistance with magnetic field, at 4K, for one of these junctions. A variation of up to 120% is obtained when the magnetization configuration of the electrodes switches from parallel (P) to antiparallel (AP). This tunnelling magnetoresistance (TMR, defined as TMR=$(R_{ap}-R_p)/R_p$, where $R_p$ and $R_{ap}$ are the resistance in the parallel and antiparallel magnetic configuration, respectively) is related to the spin-dependent density of states (DOS) of the electrodes via the Jullière formula[19]. Taking a spin-polarization of the DOS at the Fermi level SP=+90% for LSMO at the interface with STO[1,20], a TMR of 120% corresponds to a spin-polarization of +42% for conductive NFO. This value is comparable to the best results obtained with $Fe_3O_4$ electrodes[21].

The interest of the **insulating** phase of the NFO thin films is for **spin-filtering**, using the NFO film as a tunnel barrier. Because of the exchange splitting of the bands in a ferromagnet (or ferrimagnet), the height of this tunnel barrier will be different for spin-up and spin-down. Electrons tunnelling through it will experience a different barrier height depending on their spin, which will result in different transmission probabilities for spin-up and spin-down carriers. In other words, the barrier will spin-polarize the current injected even from a non-magnetic metal. If the counter-electrode has a spin-dependent DOS, a TMR effect will occur when switching the magnetic configuration of the barrier and counter-electrode from P to AP. This type of device has only been realized using the ferromagnetic semiconductor EuS as the spin-dependent barrier[22]. Insulating NFO should be a better choice, both from its much higher $T_C$ (850K vs 16K for EuS) and its larger exchange splitting (1.5-2.5 eV [23] vs 0.36 eV [24]). LSMO(35nm)/NFO(5nm)/Au and LSMO(35nm)/STO(0.8nm)NFO(5nm)/Au heterostructures (in which Au is the injector, LSMO the ferromagnetic counter-electrode, NFO the ferromagnetic insulating barrier and STO is a spacer to improve the magnetic decoupling between LSMO and NFO) have been patterned into junctions to probe this effect. Several junctions with and without STO spacers gave rise to a TMR effect. The results are qualitatively very similar for both types of structures, with slightly larger TMR for LSMO/STO/NFO structures, likely due to a better magnetic decoupling between the LSMO and NFO layers. A representative result, with a TMR of 52%, is shown in figure 4d. The corresponding spin-filtering efficiency by the NFO barrier can be estimated as 23% when one assumes a spin polarization of 90 % for LSMO [1, 20].

In summary, we have shown that oxides of spinel structure offer new opportunities for materials science. Two different phases of NiFe$_2$O$_4$ can be stabilized in nanometric epitaxial film onto perovskites. Both have a large magnetic moment, twice larger than that of the bulk, but one is conductive while the other is insulating. We have used conductive NFO layers as electrodes in magnetic tunnel junctions and shown that this material has a spin-polarization of up to 42%. We have also demonstrated that insulating NFO layers can be used as ferromagnetic tunnel barriers in a little explored type of spintronics device, spin-filter, with a spin-filtering efficiency of 23% at low temperature. These findings have two main implications. First, they evidence that NiFe$_2$O$_4$ films are interesting for spintronics as they can yield large TMR effects and they have a high Curie temperature. One may also think of combining the conductive and insulating phases of this spinel ferrite within the same heterostructure, as was done recently with the STO perovskite[25]. This would allow to build monolithic spin-filter junctions based on a single material having a large magnetic moment and high Curie temperature. Second, they open the way for the discovery of new magnetic materials, by exploiting both the many degrees of freedom of the spinel oxide structure and the epitaxial stabilization of metastable phases.

We acknowledge financial support by the Franco-Spanish PICASSO PAI program, the E.U. FP6 STREP "Nanotemplates", Contract number - NMPA4-2004-505955 and the CICyT of the Spanish government (project MAT2002-04551-cO3).

Figure captions :

Figure 1 : (a) RHEED images along [100] for the 12 nm NFO films after growth, in pure Ar or Ar/O$_2$. (b) X-ray diffraction spectra for two 12 nm NFO films close to the angular position of the (004) reflection of NFO and of the (002) reflection of STO. (c) 1 μm² AFM images of 12 nm NFO films grown in Ar/O$_2$ (main panel) and pure Ar (inset).

Figure 2 : Magnetization hysteresis cycles measured in a Quantum Design superconducting quantum interference device (SQUID) at 10K for two 6 nm NFO films. The magnetic field wa applied in plane, along the [100] direction. Insets : temperature dependence of the magnetization for a 6 nm film grown in pure Ar ; temperature dependence of the resistivity of a 12 nm NFO film grown in pure Ar. The blue symbol corresponds to the room-temperature resistivity of a 15 nm film grown in Ar/O$_2$.

Figure 3 : (a) 500 nm x 500 nm resistance mappings collected with a CTAFM for LSMO/STO(0.8nm)/NFO structures. Left : bare LSMO/STO bilayer; middle : LSMO/STO bilayer covered by 5 nm of NFO grown in pure Ar; right : LSMO/STO bilayer covered by 5 nm of NFO grown in Ar/O$_2$. (b) Dependence of the average resistance R$_{avg}$ extracted from CTAFM mappings on the NFO thickness.

Figure 4 : Structure of the LSMO/STO/NFO tunnel junctions (a) and LSMO/NFO/Au spin-filters (b), with the schematic potential profiles experienced by electrons tunnelling from top to bottom electrode. The white arrows indicate the possible orientation of the magnetization in the magnetic layers. In (b) the spin-split barrier heights inside the NFO barrier are shown in purple and blue for spin-down and spin-up electrons, respectively. (c) Magnetoresistance

R(H) cycle for a LSMO/STO/NFO magnetic tunnel junction, at 4K and a bias voltage $V_{DC}$=10mV. (d) Magnetoresistance cycle of a LSMO/STO/NFO/Au spin-filter, measured at 4K and $V_{DC}$=10mV.

Experimental

The NiFe$_2$O$_4$ films were grown by off-axis target-facing-target RF sputtering in a commercial (Plassys) UHV chamber, using two stoichiometric pressed pellets prepared by standard solid state chemistry as targets. The films were grown at 550°C and a pressure of 0.01 mbar, either of pure Ar or with 10% of O$_2$. LSMO and LSMO/STO structures were grown ex-situ by pulsed laser deposition at 700°C and an oxygen pressure of 0.46 mbar. Their surface structure as checked by RHEED before depositing the NFO film was similar to that of STO substrates. RHEED was operated at a voltage of 20 kV. X-ray diffraction was performed using Cu K$_{\alpha 1}$ radiation. The junction of figure 4c was patterned by nanoindentation lithography, and had a ~50 nm x 50 nm size. The spin-filter (12 μm² in size) of figure 4d was patterned by standard optical lithography and ion-beam etching. Magnetotransport measurements were performed with a 6kOe electromagnet in an Oxford Instruments cryostat with the field applied in plane.

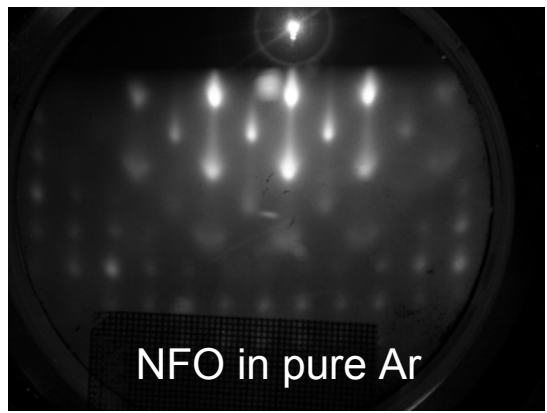 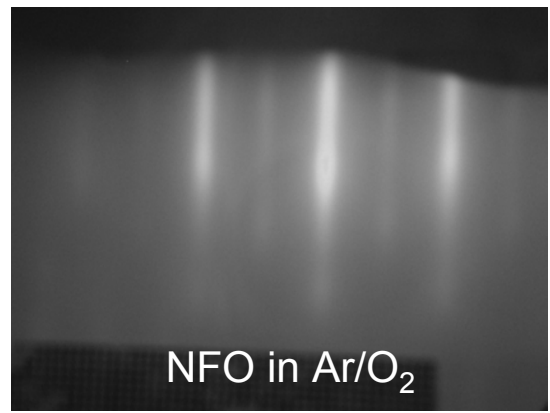

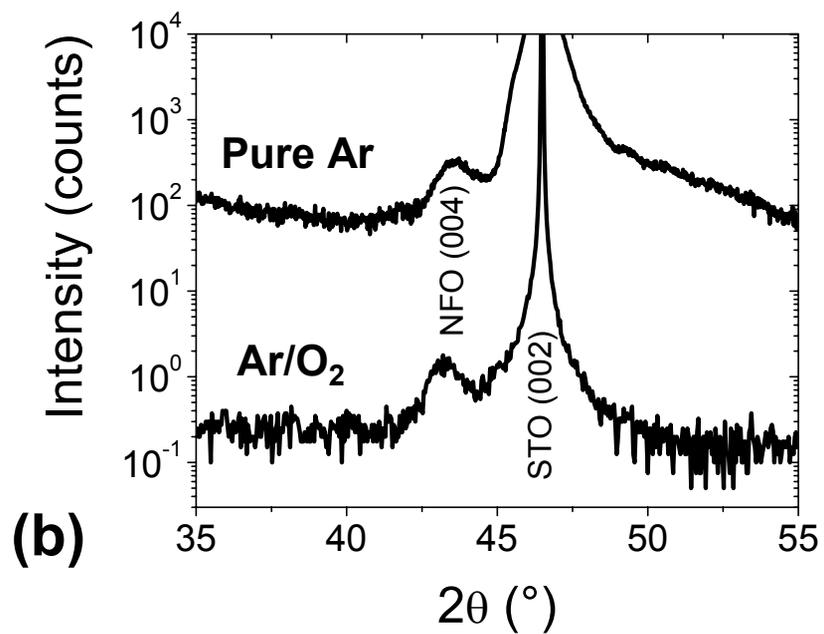 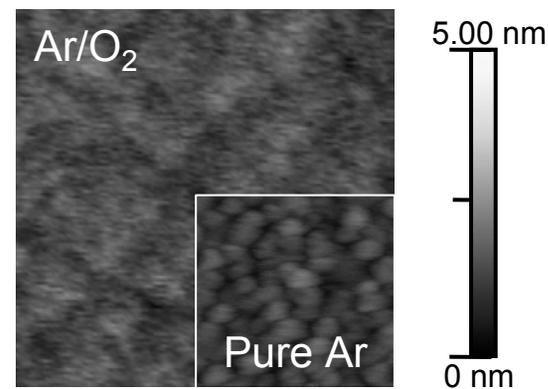

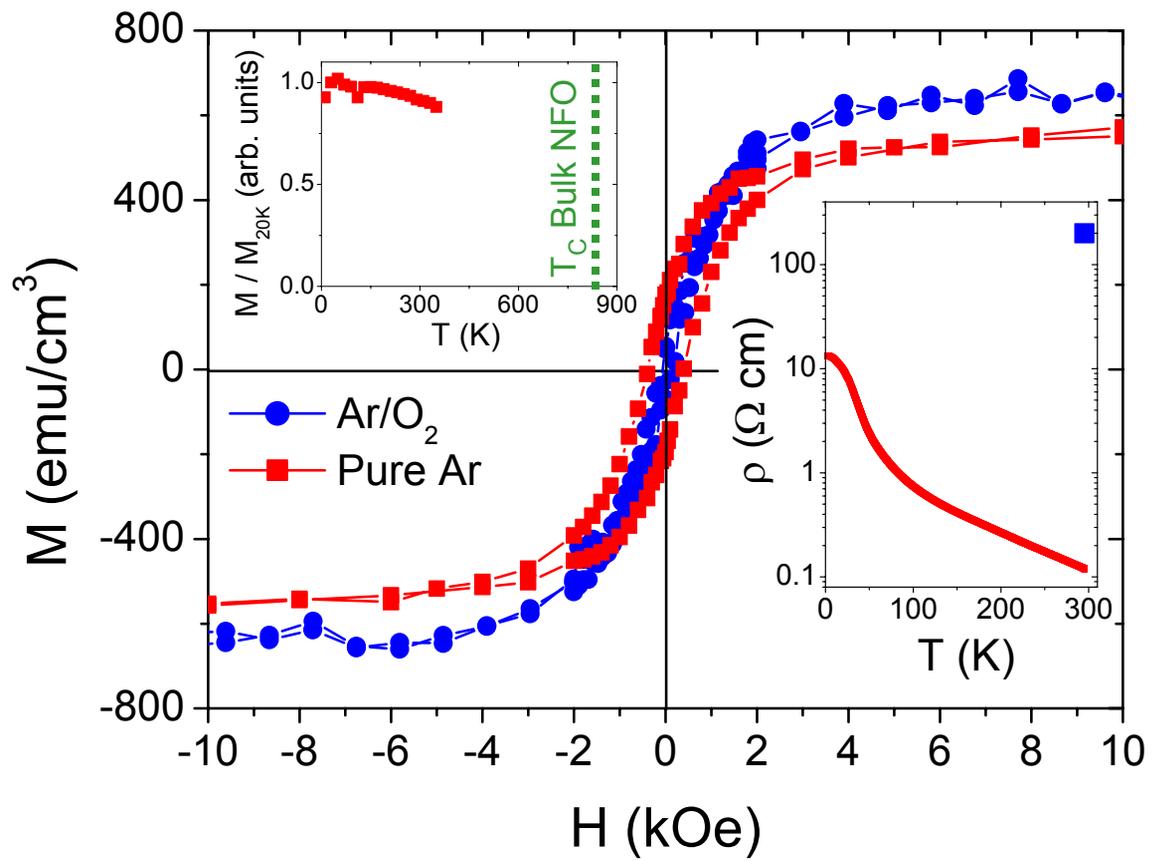

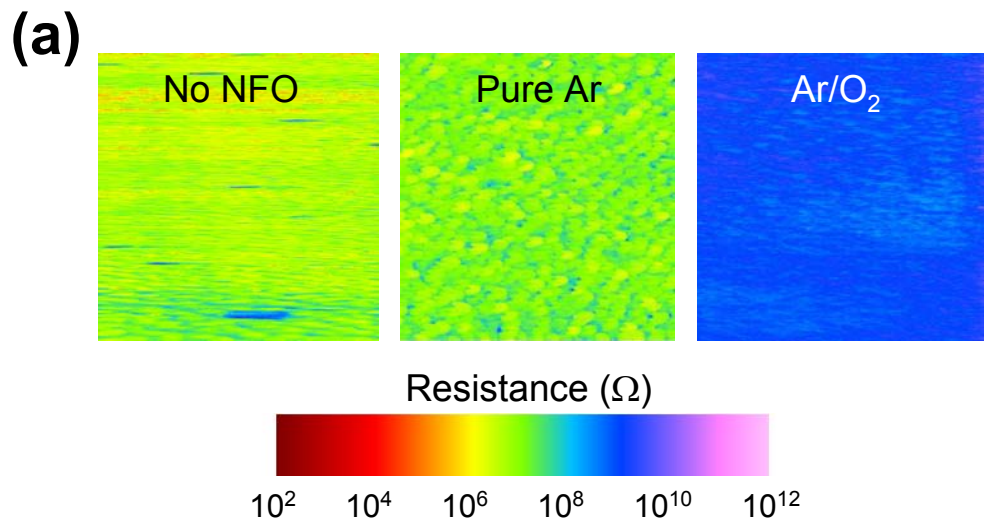
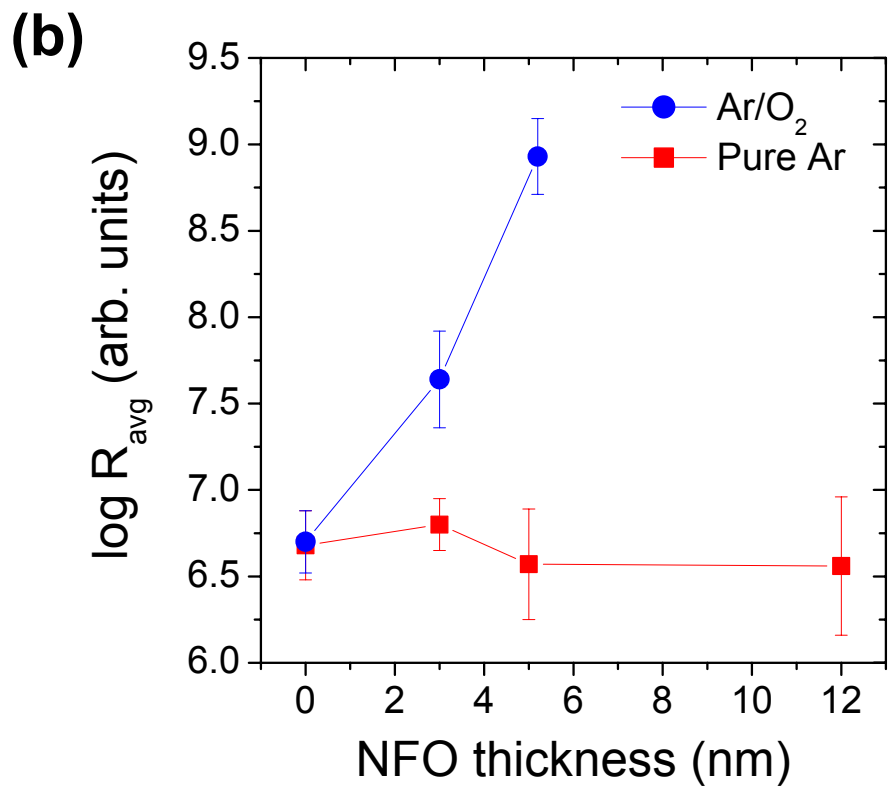

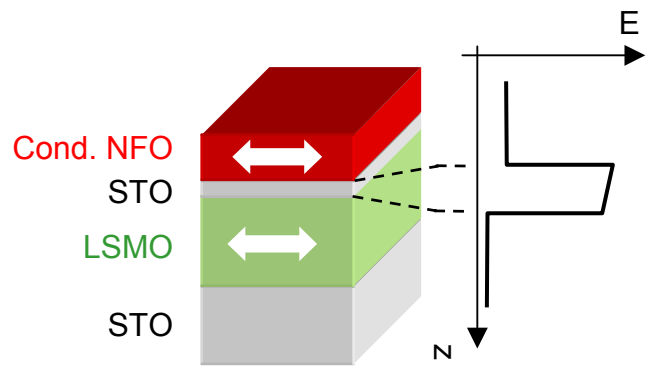 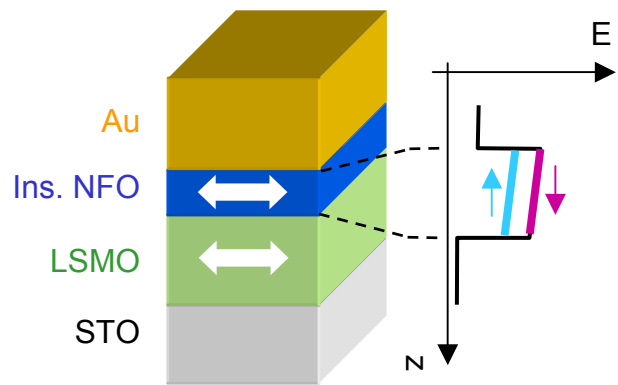
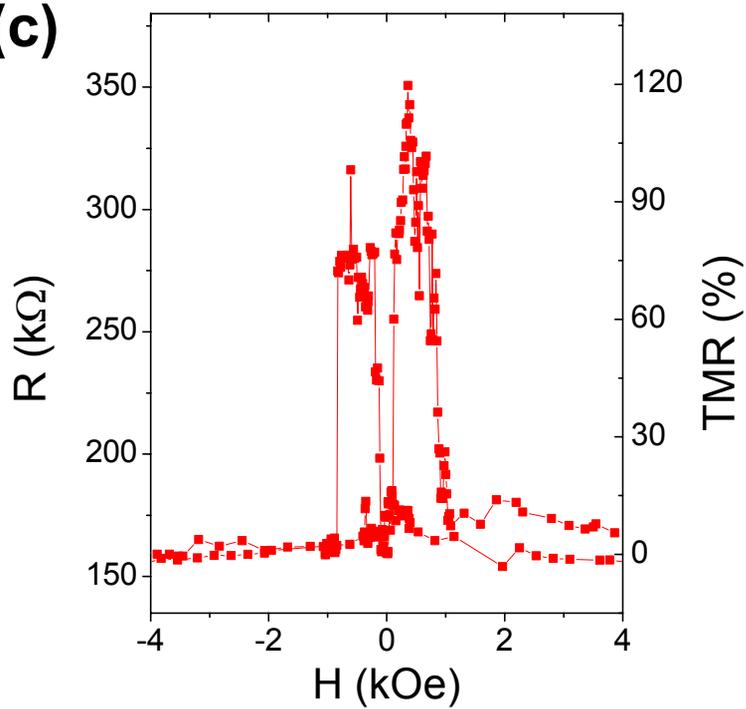 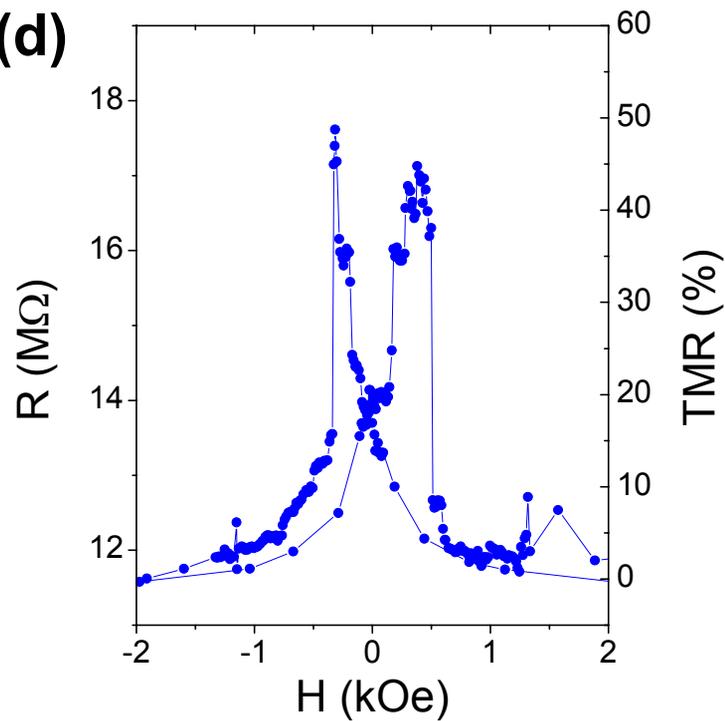